\documentclass[hidelinks,10pt,final,journal,letterpaper,twocolumn]{IEEEtran}
\ifCLASSINFOpdf
	\else
	\usepackage[dvips]{graphicx}
\fi
\usepackage[latin10]{inputenc}
\usepackage[english]{babel}
\usepackage{amsmath,amsfonts,amsthm,amssymb,mathtools}
\usepackage{algorithmic}
\usepackage{algorithm}
\usepackage{array}
\usepackage[caption=false,font=normalsize,labelfont=sf,textfont=sf]{subfig}
\usepackage{textcomp}
\usepackage{stfloats}
\usepackage{url}
\usepackage{verbatim}
\usepackage{graphicx}
\usepackage{cite}
\usepackage{url}
\usepackage{orcidlink}
\usepackage{balance}
\usepackage{xcolor}
\usepackage{ifthen}
\usepackage[acronym,nohypertypes={acronym}]{glossaries}
\usepackage{accents}
\usepackage{dsfont}
\usepackage{cancel}
\usepackage{dsfont}
\usepackage{fancyhdr}

\let\originalleft\left
\let\originalright\right
\renewcommand{\left}{\mathopen{}\mathclose\bgroup\originalleft}
\renewcommand{\right}{\aftergroup\egroup\originalright}

\newtheorem{definition}{Definition}
\newtheorem{lemma}{Lemma}

\newtheorem{proposition}{Proposition}

\newcommand{\bsf}[1]{\ensuremath{\boldsymbol{\mathsf{#1}}}}
\newcommand{\id}[1]{\mathbf{I}_{#1}}
\newcommand{\cov}[1]{\ensuremath{\mathbf{C}_{#1}}}
\DeclareMathOperator{\expec}{\mathbb{E}}
\DeclareMathOperator{\trace}{\mathrm{Tr}}
\DeclareMathOperator{\vect}{\mathrm{vec}}
\DeclareMathOperator{\prob}{\mathrm{Pr}}
\newcommand{\euler}{\ensuremath{\mathit{e}}}

\newcommand{\nr}{\ensuremath{N_{\mathrm{r}}}}
\newcommand{\nt}{\ensuremath{N_{\mathrm{t}}}}
\newcommand{\complex}{\ensuremath{\mathbb{C}}}
\newcommand{\reals}{\ensuremath{\mathbb{R}}}
\newcommand{\vecrand}[1]{\ensuremath{\widetilde{\bsf{#1}}}}
\newcommand{\vecdet}[1]{\ensuremath{\widetilde{\mathbf{#1}}}}
\newcommand{\frob}{\ensuremath{\mathrm{F}}}
\newcommand{\placeholder}{\ensuremath{\boldsymbol{\cdot}}}

\newcommand{\ie}{\textit{i.e.} }
\newcommand{\eg}{\textit{e.g.} }

\newcommand{\herm}[1]{\ensuremath{#1^{\mathrm{H}}}}
\newcommand{\trans}[1]{\ensuremath{#1^{\mathrm{T}}}}
\newcommand{\der}{\ensuremath{\mathrm{d}}}
\newcommand{\snr}{\ensuremath{\mathrm{SNR}}}
\DeclareMathOperator{\power}{\mathrm{P}}

\DeclareMathOperator*{\argmax}{arg\,max}
\DeclareMathOperator{\colsp}{\mathcal{C}}

\newacronym{csir}{CSIR}{\acrshort{csi} at the receiver}
\newacronym{ml}{ML}{maximum likelihood}
\newacronym{mimo}{MIMO}{multiple input multiple output}
\newacronym{ofdm}{OFDM}{orthogonal frequency-division multiplexing}
\newacronym{csi}{CSI}{channel state information}
\newacronym{simo}{SIMO}{single input-multiple output}
\newacronym{snr}{SNR}{signal-to-noise ratio}
\newacronym{pep}{PEP}{pairwise error probability}
\newacronym{llr}{LLR}{log-likelihood ratio}
\newacronym{asd}{ASD}{asymptotically singular detection}
\newacronym{ustm}{USTM}{unitary space-time modulation}
\newacronym{gc}{GC}{Grassmannian constellation}
\newacronym{lhs}{LHS}{left-hand side}
\newacronym{jdiv}{J-div.}{Jeffreys divergence}

\addto\extrasenglish{

}

\newcommand{\changefont}{\color{blue}\fontsize{9}{9}\selectfont}

\let\oldmaketitle\maketitle
\renewcommand{\maketitle}{%
	\oldmaketitle
	\thispagestyle{fancy}
}

\begin{document}

	\title
	{
		Singular Detection in Noncoherent Communications\thanks{This work was (partially) funded by project MAYTE (PID2022-136512OB-C21) by MICIU/AEI/10.13039/501100011033 and ERDF/EU, grant 2021 SGR 01033 and grant 2022 FI SDUR 00164 by Departament de Recerca i Universitats de la Generalitat de Catalunya.}%
		\thanks{The authors are with the Signal Processing and Communications Group (SPCOM), Departament de Teoria del Senyal i Comunicacions (TSC), Universitat Politècnica de Catalunya (UPC), 08034 Barcelona, Spain (e-mail: \{marc.vila.insa, jaume.riba\}@upc.edu).}
	}
	
	\author{Marc Vil\`{a}-Insa \orcidlink{0000-0002-7032-1411}, \IEEEmembership{Graduate Student Member, IEEE}, and Jaume Riba \orcidlink{0000-0002-5515-8169}, \IEEEmembership{Senior Member, IEEE}}
	
	\markboth{}{}
	
	\maketitle
	
	\begin{abstract}
		This paper proposes a general analysis of codeword detection in noncoherent communications.
		Motivated by the existence of error floors in various regimes, fundamental characteristics of signal design are investigated.
		In particular, the necessary and sufficient conditions for asymptotically singular detection (\ie error-free in the limit) are derived from classical results in detection theory.
		By leveraging tools from linear algebra and the theory of Hilbert spaces, we are able to characterize asymptotic singularity in two main scenarios: the large array and high \acrshort{snr} regimes.
		The results generalize previous works and extend the notion of unique identification, as well as re-contextualize the geometry of Grassmannian constellations from an alternative perspective.
	\end{abstract}
	
	\begin{IEEEkeywords}
		Noncoherent communications, singular detection, unique identification, massive \acrshort{mimo}, high \acrshort{snr} regime.
	\end{IEEEkeywords}
	
	\section{Introduction}\label{sec:intro}
	
		\IEEEPARstart{N}{oncoherent} detection has been a part of cellular wireless communications since their inception~\cite[App.~D]{Goldsmith2005}.
		First and second generation systems were defined by device manufacturing limitations, which constrained communication schemes to employ modulations that did not require instantaneous \acrfull{csi}~\cite{Chowdhury2016}.
		With technological advancements came a rise in demand for higher data-rates, making spectral resources more valuable.
		Therefore, digital systems based on coherent detection became the norm in subsequent generations (\ie 3G to 5G), due to their improved spectral efficiency.
		
		New applications emerging within the fifth and succeeding generations showcase novel technical bottlenecks beyond spectrum scarcity.
		Several of these barriers are related to the acquisition of reliable instantaneous \acrshort{csi}, especially when employing large numbers of antennas, in high mobility scenarios or under low latency requirements~\cite{Chafii2023}.
		All these challenges have rekindled an interest in noncoherent solutions for next generation communications.
		
		The present work studies the error performance of noncoherent systems from the perspective of detection theory.
		Motivated by the existence of error floors in various communication settings~\cite{Jing2016,VilaInsa2024a}, we analyze which signal properties allow for an \textit{\acrfull{asd}}, \ie asymptotically error-free.
		Understanding if a configuration will display a fundamental error floor provides valuable insights on the achievable gains obtained by pouring more resources into a system.
		In particular, we explore the following scenarios:
		\begin{itemize}
			\item \textbf{Large array regime} ($\text{no. receiver antennas}\to\infty$).
			It sheds light onto the performance improvements brought by massive arrays (such as the emergence of \textit{channel hardening}~\cite{Jing2016}), against the infrastructure costs they entail.
			\item \textbf{High \acrfull{snr} regime} ($\snr\to\infty$).
			Likewise, this analysis is of particular interest to assess the gains attainable by increasing the transmitted power, in view of energy efficiency and power-limited systems~\cite{Chafii2023}.
		\end{itemize}
		
		The main goal behind this work is to establish necessary and sufficient conditions for an alphabet to yield \acrshort{asd} under the presented regimes.
		This allows to understand fundamental limitations of noncoherent systems and unveils integral aspects of codeword design.
		Therefore, without sacrificing interpretability, we have considered a signal framework that encapsulates a variety of scenarios of interest, both
		well-established~\cite[Sec.~3.6.1]{Heath2018} and state-of-the-art~\cite{Pizzo2022}.
		The main results are summarized next.
		In Section~\ref{sec:nr}, we take some classical ideas from detection theory~\cite{Kailath1975} and adapt them for the large array regime (\textit{Proposition~\ref{prop:nr}}), with which we determine novel insightful requirements for \acrshort{asd} in such scenario.
		Conversely, in Section~\ref{sec:high_snr} we derive the necessary and sufficient conditions for \acrshort{asd} in the high \acrshort{snr} regime (\textit{Proposition~\ref{prop:subspace}}), which yield a powerful geometric interpretation on the problem.
		These results are a refinement of Theorems 1 and 2 from~\cite{VilaInsa2024a}, and generalize them for any configuration of transmitting and receiving antennas and codeword length.
		
		The notation used throughout the text is defined next.
		Vectors and matrices: boldface lowercase and uppercase.
		Transpose and conjugate transpose: $\trans{\placeholder}$, $\herm{\placeholder}$.
		Trace: $\trace\{\placeholder\}$.
		Minimum and maximum eigenvalues: $\sigma_{\min}(\placeholder)$ and $\sigma_{\max}(\placeholder)$.
		Entry ($r$,$c$) of a matrix: $[\mathbf{A}]_{r,c}$.
		Column space: $\colsp(\placeholder)$.
		Column-wise vectorization: $\vect(\placeholder)$.
		Kronecker product: $\otimes$.
		Matrix determinant: $\lvert\mathbf{A}\rvert$.
		Euclidean, Frobenius and weighted norms: $\lVert\mathbf{a}\rVert$, $\lVert\mathbf{X}\rVert_{\frob}$, $\lVert\mathbf{X}\rVert_{\mathbf{A}}\triangleq\sqrt{\trace\{\herm{\mathbf{X}}\mathbf{A}^{-1}\mathbf{X}\}}$.
		Empty set: $\{\emptyset\}$.
		Direct sum: $\oplus$.
		Random variables: sans serif font.
		Expectation: $\expec[\placeholder]$.
		Circularly symmetric complex normal vector: $\bsf{a}\sim\mathcal{CN}(\mathbf{m},\mathbf{C})$.
	
	\section{Preliminary notions}
	
		\subsection{Signal model}\label{ssec:model}
	
			Consider a \acrshort{mimo} point-to-point system, in which transmitter and receiver are equipped with $\nt$ and $\nr $ antennas, respectively.
			The channel is assumed frequency flat with a coherence time $K$.
			This translates into a channel matrix $\bsf{H}\in\complex^{\nt\times\nr}$ that remains constant for $K$ channel uses (\ie block flat fading).
			During each block, the transmitter sends an equiprobable codeword $\mathbf{S}\in\complex^{K\times\nt}$ selected from a finite alphabet $\mathcal{S}$ of size $M$.
			We assume an average power constraint of $\frac{1}{K}\expec_{\bsf{S}}[\lVert\bsf{S}\rVert_{\mathbf{\frob}}^2]\triangleq1$.
			
			The signal at the receiver is expressed as a time-space matrix, using a complex baseband representation:
			\begin{equation}
				\bsf{Y}=\bsf{X}\bsf{H}+\bsf{Z}\in\complex^{K\times\nr},\label{eq:model}
			\end{equation}
			where $\bsf{X}\triangleq\sqrt{\power_{\bsf{X}}}\cdot\bsf{S}$ is the transmitted signal with average power $\power_{\bsf{X}}$, and $\bsf{Z}$ is an independent additive Gaussian noise component.
			The average channel and noise power is normalized as
			\begin{equation}
				\expec_{\bsf{H}}[\lVert\bsf{H}\rVert_{\frob}^2] \triangleq 1, \quad 	\tfrac{1}{K}\expec_{\bsf{Z}}[\lVert\bsf{Z}\rVert_{\frob}^2] \triangleq \power_{\bsf{Z}}, \label{eq:energy_const}
			\end{equation}
			with which we define the \acrshort{snr} at the receiver:
			\begin{equation}
				\snr \triangleq 		\frac{\expec_{\bsf{H},\bsf{X}}[\lVert\bsf{XH}\rVert_{\frob}^2]}{\expec_{\bsf{Z}}[\lVert\bsf{Z}\rVert_{\frob}^2]} = \frac{\power_{\bsf{X}}}{\power_{\bsf{Z}}} \cdot \frac{\expec_{\bsf{H},\bsf{S}}[\lVert\bsf{SH}\rVert_{\frob}^2]}{K}.\label{eq:snr}
			\end{equation}
			Under this model, the received signal will be finite-energy regardless of $\nr$, which is relevant in the study of the large array regime in Section~\ref{sec:nr}.
			Indeed, from the power normalizations in~\eqref{eq:energy_const}, the average received energy when transmitting $\mathbf{S}$ is
			\begin{equation}
				\expec_{\bsf{Y}\vert\mathbf{S}}[\lVert\bsf{Y}\rVert_{\frob}^2] = \power_{\bsf{X}}\expec_{\bsf{H}}[\lVert\mathbf{S}\bsf{H}\rVert_{\frob}^2] + \power_{\bsf{Z}}K.
			\end{equation}
			Applying~\cite[Fact~10.14.22]{Bernstein2018} and the maximum transmitted codeword energy,
			\begin{equation}
				\expec_{\bsf{Y}\vert\mathbf{S}}[\lVert\bsf{Y}\rVert_{\frob}^2] \leq \power_{\bsf{X}}MK + \power_{\bsf{Z}}K, \label{eq:ineq}
			\end{equation}
			which is clearly bounded for any $\nr$.
			
			The distributions of both $\bsf{H}$ and $\bsf{Z}$ are assumed to be known by the receiver but not their realizations\footnote{This is referred to as \textit{statistical \acrlong{csir}}.}.
			To characterize them statistically, it is convenient to vectorize the received signal matrix column-wise as follows:
			\begin{equation}
				\begin{aligned}
					\vecrand{y}\triangleq\vect(\bsf{Y})&=(\id{\nr}\otimes\bsf{X})\vect(\bsf{H})+\vect(\bsf{Z})\\
					&\triangleq\vecrand{X}\vecrand{h}+\vecrand{z}\in\complex^{K\nr }.
				\end{aligned}\label{eq:vec}
			\end{equation}
			We define $\vecrand{S}\triangleq\id{\nr}\otimes\bsf{S}$ in the same manner.
			Assuming correlated Rayleigh fading, the vectorized channel matrix is distributed as $\vecrand{h}\sim\mathcal{CN}(\mathbf{0}_{\nt\nr },\cov{\vecrand{h}})$.
			Similarly, the noise is distributed as $\vecrand{z}\sim\mathcal{CN}(\mathbf{0}_{K\nr },\cov{\vecrand{z}})$ and its covariance matrix is assumed full-rank, without loss of generality.
	
		\subsection{Error probability of ML detection}\label{ssec:error}
			
			An important metric to consider when designing a digital communication system is the error probability of codeword detection.
			When dealing with an equiprobable alphabet,	the receiver that minimizes it is the \textit{\acrfull{ml} detector}~\cite[Thm.~21.3.3]{Lapidoth2017}.
			It is derived from the likelihood function of the received signal~\eqref{eq:vec}, conditioned to a transmitted codeword $\mathbf{S}_i$ (\ie $\mathbf{X}_i$) and a channel realization $\mathbf{H}$:
			\begin{equation}
				\mathrm{f}_{\bsf{Y}\vert\mathbf{S}_i,\mathbf{H}}(\vecdet{y})=\tfrac{1}{\pi^{K\nr}\lvert\cov{\vecrand{z}}\rvert}\euler^{-\herm{(\vecdet{y}-\vecdet{X}_i\vecdet{h})}\cov{\vecrand{z}}^{-1}(\vecdet{y}-\vecdet{X}_i\vecdet{h})}.
			\end{equation}
			Since the channel realization is unknown at the receiver, the uncertainty of $\bsf{H}$ can be treated as a random variable (\ie \textit{unconditional model}~\cite{PetreStoica1990}) and removed from the conditioning by marginalization:
			\begin{equation}
				\mathrm{f}_{\bsf{Y}\vert\mathbf{S}_i}(\vecdet{y})=\expec_{\bsf{H}}[\mathrm{f}_{\bsf{Y}\vert\mathbf{S}_i,\mathbf{H}}(\vecdet{y})]=\tfrac{1}{\pi^{K\nr}\lvert\cov{i}\rvert}\euler^{-\herm{\vecdet{y}}\cov{i}^{-1}\vecdet{y}},\label{eq:likelihood}
			\end{equation}
			where
			$\cov{i}\triangleq\vecdet{X}_i\cov{\vecrand{h}}\vecdet{X}\herm{\vphantom{\mathbf{X}}}_i+\cov{\vecrand{z}}$
			is the covariance matrix of the received signal conditioned to $\mathbf{S}_i$.
			With~\eqref{eq:likelihood} we can obtain the \acrshort{ml} detector by maximizing it over all possible codewords from $\mathcal{S}$:
			$\widehat{\mathbf{S}}_{\mathrm{ML}}=\argmax_{\mathbf{S}_i\in\mathcal{S}}\mathrm{f}_{\bsf{Y}\vert\mathbf{S}_i}(\vecdet{y})$.

		\subsection{Pairwise error probability}
		
			The probability of error of a communication system is usually analytically intractable, so various works in the literature~\cite{Ngo2022} resort to the much simpler \textit{\acrfull{pep}} between two codewords.
			It is defined as:
			\begin{equation}
				\begin{aligned}
					\mathrm{P}_{a\to b}&\triangleq\prob\{\mathrm{f}_{\bsf{Y}\vert\mathbf{S}_a}(\vecrand{y})\leq \mathrm{f}_{\bsf{Y}\vert\mathbf{S}_b}(\vecrand{y}) \,\vert\, \bsf{S}=\mathbf{S}_a\}\\
					&=\prob\{\mathrm{L}_{a,b}(\vecrand{y})\leq0 \,\vert\, \bsf{S}=\mathbf{S}_a\},
				\end{aligned}\label{eq:pep}
			\end{equation}
			where
			\begin{equation}
				\mathrm{L}_{a,b}(\vecdet{y})\triangleq\ln\frac{\mathrm{f}_{\bsf{Y}\vert\mathbf{S}_a}(\vecdet{y})}{\mathrm{f}_{\bsf{Y}\vert\mathbf{S}_b}(\vecdet{y})}=\herm{\vecdet{y}}(\cov{b}^{-1}-\cov{a}^{-1})\vecdet{y}-\ln\frac{\lvert\cov{a}\rvert}{\lvert\cov{b}\rvert}\label{eq:llr}
			\end{equation}
			is the \textit{\acrfull{llr}} between hypotheses $a$ and $b$.
			
			The \acrshort{pep}s of an alphabet are very insightful tools to analyze the performance of its design.
			With the maximum \acrshort{pep} of $\mathcal{S}$, we can bound its detection error probability $\mathrm{P}_{\mathrm{e}}$ as~\cite{Ngo2022}:
			\begin{equation}
				\max_{\mathbf{S}_a\neq\mathbf{S}_b\in\mathcal{S}}\tfrac{1}{M}\mathrm{P}_{a\to b}\leq\mathrm{P}_{\mathrm{e}}\leq\max_{\mathbf{S}_a\neq\mathbf{S}_b\in\mathcal{S}}(M-1)\mathrm{P}_{a\to b}.
			\end{equation}
			This implies the detection error probability of $\mathcal{S}$ will vanish if and only if its maximum \acrshort{pep} does as well.
			
		\subsection{Unique identification}
				
			\begin{definition}\label{def:ui}
				An alphabet $\mathcal{S}$ is {\normalfont uniquely identifiable}~\cite{VilaInsa2024a} if
				\begin{equation}
					\cov{a}\neq\cov{b}\iff\mathbf{S}_a\neq\mathbf{S}_b,\quad\forall\mathbf{S}_a,\mathbf{S}_b\in\mathcal{S}.
				\end{equation}
			\end{definition}
			This property is fundamental to communication systems working noncoherently, as displayed in the following lemma.
			\begin{lemma}\label{lmm:ui}
				Unique identification is a \textit{necessary condition} for an alphabet to be detected with arbitrarily low $\mathrm{P}_{\mathrm{e}}$ under the model from Section~\ref{ssec:model}.
			\end{lemma}
			\begin{IEEEproof}
				The proof is straightforward.
				Having $\cov{a}=\cov{b}$ for two different codewords $\mathbf{S}_a,\mathbf{S}_b\in\mathcal{S}$ collapses their \acrshort{llr}, \ie $\mathrm{L}_{a,b}(\vecdet{y})=0$, $\forall\vecdet{y}\in\complex^{K\nr}$.
				This prevents the receiver from distinguishing between them, so their associated \acrshort{pep} is always positively lower-bounded.
				Therefore, unique identification is necessary for singular detection.
			\end{IEEEproof}
		
	\section{Large array regime}\label{sec:nr}
	
		The next result is a restatement of~\cite[Lemma~3]{Kailath1975}, which is based on~\cite[Ch.~8,~9]{Hajek1998}.
		We consider it to establish under which criteria noncoherent detection will benefit from increasing $\nr$.
		To expound it, we introduce the \textit{\acrfull{jdiv}}\footnote{Some authors (see~\cite{Wang2005}) define the \acrshort{jdiv} as half of~\eqref{eq:jeffreys}.}~\cite[Ch.~8]{Hajek1998} between the distributions of $\vecrand{y}\vert\mathbf{S}_a$ and $\vecrand{y}\vert\mathbf{S}_b$:
		\begin{equation}
			\mathrm{J}_{\nr} \triangleq \expec_{\bsf{Y}\vert\mathbf{S}_a}[\mathrm{L}_{a,b}(\vecrand{y})]-\expec_{\bsf{Y}\vert\mathbf{S}_b}[\mathrm{L}_{a,b}(\vecrand{y})]. \label{eq:jeffreys}
		\end{equation}
		Taking this into account, the aforementioned result is displayed next.
		\begin{proposition}[Kailath \& Weinert~\cite{Kailath1975}]\label{prop:nr}
			A {\normalfont necessary} and {\normalfont sufficient} condition for alphabet $\mathcal{S}$ to be detected with vanishing $\mathrm{P}_{\mathrm{e}}$ as $\nr\to\infty$ under the model from Section~\ref{ssec:model} is
			\begin{equation}
				\lim_{\nr\to\infty}\mathrm{J}_{\nr} = \infty,
			\end{equation}
			for all $\mathbf{S}_a\neq\mathbf{S}_b$ in $\mathcal{S}$.
		\end{proposition}
		
		Particularizing the \acrshort{jdiv} to our signal model yields
		\begin{equation}
			\mathrm{J}_{\nr} = \trace\{(\cov{b}^{-1}-\cov{a}^{-1})(\cov{a} - \cov{b})\},\label{eq:j-div}
		\end{equation}
		where we have used the circularity and linearity properties of the trace.
		The signal normalization considered herein allows the obtained criterion to only capture nontrivial situations~\cite{Kailath1975}.
		On the one hand, singularity will not emerge for finite $\nr$ thanks to\footnote{The symbol $``\diamond"$ will be used throughout the text as a placeholder for both $``a"$ and $``b"$ indistinctly.} $\cov{\diamond}$ being strictly positive-definite, due to the rank-completeness of $\cov{\vecrand{z}}$.
		On the other hand, \acrshort{asd} will never arise as a result of trivially increasing the received power, since all the entries of $\cov{\diamond}$ are bounded.
		This can be shown with the Cauchy-Schwarz inequality:
		\begin{equation}
			\textstyle \lvert[\cov{\diamond}]_{r,c}\rvert \leq \sqrt{[\cov{\diamond}]_{r,r} [\cov{\diamond}]_{c,c}} \leq \trace\{\cov{\diamond}\} < \infty.
		\end{equation}
		The last inequality can be stated because $\vecrand{y}\vert\mathbf{S}_{\diamond}$ is finite-energy for any $\nr$, as proved in~\eqref{eq:ineq}.
		
		We may express~\eqref{eq:j-div} in the more insightful form
		\begin{equation}
			\mathrm{J}_{\nr} = \bigl\lVert\cov{a}^{-\frac{1}{2}}(\cov{a}-\cov{b})\cov{b}^{-\frac{1}{2}}\bigr\rVert_{\frob}^2.\label{eq:jdiv_norm}
		\end{equation}
		With it, Proposition~\ref{prop:nr} defines a notion of unique identification for increasing dimensionality: the inequality from Lemma~\ref{lmm:ui} is replaced by a divergent norm, in a similar manner as how equality and strong convergence are related when dealing with Hilbert spaces~\cite[Sec.~2.8.1]{Kennedy2013}.
		
		In~\cite[Thm.~1]{VilaInsa2024a}, a criterion for \acrshort{asd} in the large array regime was derived for a single channel use ($K=1$) \acrshort{simo} ($\nt=1$) system.
		Not only is Proposition~\ref{prop:nr} a more general result applicable to any configuration of $K$ and $\nt$, but it also entails a clear refinement.
		Whereas in~\cite{VilaInsa2024a} elaborate mathematical machinery had to be deployed (\eg \textit{Cantelli's inequality} and convergence tests), Proposition~\ref{prop:nr} allows us to reach the conclusions presented in that particular analysis from a much more straightforward procedure.
		Indeed, particularizing~\eqref{eq:jdiv_norm} for $K=\nt=1$ yields
		\begin{equation}
			\mathrm{J}_{\nr}=\Delta_{a,b}^2\cdot \trace\{\cov{\vecrand{h}}\cov{b}^{-1}\cov{\vecrand{h}}\cov{a}^{-1}\}, \label{eq:crit}
		\end{equation}
		for $\Delta_{a,b}\triangleq\lvert x_a\rvert^2 - \lvert x_b\rvert^2$.
		With simple manipulations we can bound the previous expression as
		\begin{equation}
			\tfrac{\Delta_{a,b}^2}{(\lvert x_a\rvert^2C+ 1)(\lvert x_b\rvert^2C + 1)}\cdot\trace\{\mathbf{\Gamma}^2\}\leq\mathrm{J}_{\nr}\leq\Delta_{a,b}^2\cdot\trace\{\mathbf{\Gamma}^2\}, \label{eq:bounds}
		\end{equation}
		where $\mathbf{\Gamma}\triangleq\cov{\vecrand{z}}^{-\frac{1}{2}}\cov{\vecrand{h}}\cov{\vecrand{z}}^{-\frac{1}{2}}$ and $C\triangleq\sigma_{\max}(\mathbf{\Gamma})>0$.
		This allows splitting the divergence study of~\eqref{eq:crit} into two simpler criteria.
		The first one depends on the statistics of channel and noise in the studied model, which must yield $\lim_{\nr\to\infty}\trace\{\mathbf{\Gamma}^2\}=\infty$.
		This is related to the decay rate of signal and noise spectra discussed in~\cite[Sec.~3]{Kailath1998}.
		If this condition is met, the second one depends on the design of the transmitted signal: \acrshort{asd} will be achieved for $\lvert x_a\rvert^2 \neq \lvert x_b\rvert^2$, so that $\Delta_{a,b}^2$ in~\eqref{eq:crit} and~\eqref{eq:bounds} is strictly positive.
		
		The \acrshort{jdiv} measures how different $\mathrm{f}_{\bsf{Y}\vert\mathbf{S}_a}$ and $\mathrm{f}_{\bsf{Y}\vert\mathbf{S}_b}$ are, and is null if and only if $\mathbf{S}_a$ and $\mathbf{S}_b$ are not uniquely identifiable~\cite[Prop.~12]{Moakher2012}.
		Moreover, when $\vecrand{y}\vert\mathbf{S}_a$ and $\vecrand{y}\vert\mathbf{S}_b$ yield close distributions, the \acrshort{jdiv} approximates the squared geodesic distance between them over the statistical manifold defined by the Fisher information metric~\cite{Wang2005}.
		This distance is commonly expressed in terms of the \textit{normalized covariance matrix}~\cite{Moakher2012} $\mathring{\mathbf{C}}\triangleq\cov{b}^{-\frac{1}{2}}\cov{a}\cov{b}^{-\frac{1}{2}}$.
		Similarly, we can use it to represent $\mathrm{J}_{\nr}$:
		\begin{equation}
			\mathrm{J}_{\nr} = \lVert\mathring{\mathbf{C}}-\id{K\nr}\rVert_{\mathring{\mathbf{C}}}^2, \label{eq:j_normalized}
		\end{equation}
		which clearly measures how much $\mathring{\mathbf{C}}$ differs from $\id{K\nr}$.
		This dissimilarity (and thus the one between $\cov{a}$ and $\cov{b}$) must increase with $\nr$ for $\mathrm{J}_{\nr}$ to diverge.
			
		On a final note, a set of equivalent conditions for \acrshort{asd} can be derived from~\eqref{eq:j_normalized}, as acknowledged in~\cite[Lemma 4]{Kailath1975}.
		The \acrshort{jdiv} will diverge if and only if
		\begin{equation}
			\lVert\mathring{\mathbf{C}}-\id{K\nr}\rVert_{\frob}^2 = \lVert\cov{a}-\cov{b}\rVert_{\cov{b}}^2 \to \infty,\quad\sigma_{\min}(\mathring{\mathbf{C}})>0, \label{eq:j_simple}
		\end{equation}
		as $\nr\to\infty$.
		One would intuitively expect the metric that arises for \acrshort{asd} in the large array regime to be $\lVert\cov{a}-\cov{b}\rVert_{\frob}^2$ (\ie the Euclidean distance), as an extension of Lemma~\ref{lmm:ui}.
		Remarkably, we have established in~\eqref{eq:jdiv_norm} that this is not the case.
		Even with the resemblance between~\eqref{eq:j_simple} and the intuitive Euclidean distance, the relevant metric to assess \acrshort{asd} is a norm defined from hypotheses $a$ and $b$~\cite{Kailath1975}.

	\section{High SNR regime}\label{sec:high_snr}
		
		The next results are derived for full-rank $\cov{\vecrand{h}}$.
		Their extension to the rank-deficient case can be obtained by transforming the model in Section~\ref{ssec:model} onto a lower-dimensional space.
		
		\begin{proposition}\label{prop:subspace}
			A {\normalfont necessary} and {\normalfont sufficient} condition for a constellation $\mathcal{S}$ to be detected with vanishing $\mathrm{P}_{\mathrm{e}}$ as $\snr\to\infty$ under the model from Section~\ref{ssec:model} is
			\begin{equation}
				\colsp(\mathbf{S}_a) \neq \colsp(\mathbf{S}_b) \iff \mathbf{S}_a\neq\mathbf{S}_b, \quad\forall \mathbf{S}_a,\mathbf{S}_b\in\mathcal{S}.\label{eq:prop_condition_2}
			\end{equation}
		\end{proposition}
		\begin{IEEEproof}
			Let $\gamma\triangleq\power_{\bsf{X}}/\power_{\bsf{Z}}$.
			We can restate definition~\eqref{eq:snr} as $\snr = \gamma \cdot \expec_{\bsf{S}}[\trace\{\vecrand{S}\herm{\vphantom{\bsf{S}}}\vecrand{S}\cov{\vecrand{h}}\}]/K$.
			Using~\cite[Prop.~10.4.13]{Bernstein2018}, we can bound the \acrshort{snr} as
			\begin{equation}
				\gamma\nr\cdot\sigma_{\min}(\cov{\vecrand{h}}) \leq \snr \leq \gamma\nr\cdot\sigma_{\max}(\cov{\vecrand{h}}).
			\end{equation}
			Therefore, $\snr\to\infty \iff \gamma\to\infty$.
			
			The \acrshort{pep} in~\eqref{eq:pep} can be expressed in integral form:
			\begin{equation}
				\mathrm{P}_{a\to b}=\prob\{\vecrand{y}\in\mathcal{P}\vert\bsf{S}=\mathbf{S}_a\}=\int_{\mathcal{P}}\mathrm{f}_{\bsf{Y}\vert\mathbf{S}_a}(\vecdet{y})\der\vecdet{y},\label{eq:pep_integral}
			\end{equation}
			as stated in~\cite[Sec.~20.5]{Lapidoth2017}.
			The region of integration is
			\begin{equation}
				\mathcal{P}\triangleq\{\vecdet{y}\in\complex^{K\nr}:\mathrm{L}_{a,b}(\vecdet{y})\leq 0\}.
			\end{equation}
			We are interested in the behavior of~\eqref{eq:pep_integral} as $\gamma\to\infty$.
			As a first step, we define the normalized covariance matrix $\overline{\mathbf{C}}_{\vecrand{z}}\triangleq\frac{1}{\power_{\bsf{Z}}}\cov{\vecrand{z}}$, with which we whiten the received signal: $\mathbf{r}\triangleq (\power_{\bsf{X}}\overline{\mathbf{C}}_{\vecrand{z}})^{-\frac{1}{2}}\vecdet{y}$.
			The \acrshort{pep} integral in terms of $\mathbf{r}$ becomes~\cite[Lemma~17.4.6]{Lapidoth2017}:
			\begin{equation}
				\mathrm{P}_{a\to b} = \int_{\mathcal{P}'} \tfrac{1}{\pi^{K\nr}\lvert\mathbf{D}_a(\gamma)\rvert} \euler^{-\herm{\mathbf{r}}\mathbf{D}_{a}^{-1}(\gamma)\mathbf{r}} \der\mathbf{r}, \label{eq:pep_snr}
			\end{equation}
			where we have defined
			\begin{equation}
				\mathbf{D}_{\diamond}(\gamma)\triangleq\mathbf{\Xi}_{\diamond}+\tfrac{1}{\gamma}\id{K\nr},\quad\mathbf{\Xi}_{\diamond}\triangleq\overline{\mathbf{C}}_{\vecrand{z}}^{-\frac{1}{2}}\vecdet{S}_{\diamond}\cov{\vecrand{h}}\vecdet{S}\herm{\vphantom{\mathbf{S}}}_{\diamond}\overline{\mathbf{C}}_{\vecrand{z}}^{-\frac{1}{2}}.\label{eq:D}
			\end{equation}
			The new region of integration is
			\begin{equation}
				\mathcal{P}'=\Bigl\{\mathbf{r}\in\complex^{K\nr}:\lVert\mathbf{r}\rVert_{\mathbf{D}_b(\gamma)}^2-\lVert\mathbf{r}\rVert_{\mathbf{D}_a(\gamma)}^2\leq\ln\tfrac{\lvert\mathbf{D}_a(\gamma)\rvert}{\lvert\mathbf{D}_b(\gamma)\rvert}\Bigr\}.
			\end{equation}
			
			To analyze the limit of the \acrshort{pep} when $\gamma\to\infty$, it is convenient to restate~\eqref{eq:pep_snr} in terms of the following orthogonal decomposition: $\complex^{K\nr} \triangleq \mathrm{V}_{\diamond}\oplus\mathrm{V}_{\diamond}^{\bot}$.
			The $N_{\diamond}$-dimensional signal subspace $\mathrm{V}_{\diamond}$ is spanned by $\mathbf{U}_{\diamond}\in\complex^{K\nr\times N_{\diamond}}$, which contains the eigenvectors of $\mathbf{\Xi}_{\diamond}$.
			Its non-zero eigenvalues are positive and can be grouped in a diagonal matrix $\mathbf{\Lambda}_{\diamond}\in\reals^{N_{\diamond}\times N_{\diamond}}$.
			Similarly, $\mathrm{V}_{\diamond}^{\bot}$ is the orthogonal complement of $\mathrm{V}_{\diamond}$, which corresponds to the noise subspace and has dimension $N_{\diamond}^{\bot}\triangleq K\nr-N_{\diamond}$.
			It is spanned by $\mathbf{U}_{\diamond}^{\bot}\in\complex^{K\nr\times N_{\diamond}^{\bot}}$.
			
			With the previous definitions,~\eqref{eq:D} becomes
			\begin{equation}
				\mathbf{D}_{\diamond}(\gamma) \triangleq \mathbf{U}_{\diamond}\bigl(\mathbf{\Lambda}_{\diamond} + \tfrac{1}{\gamma}\id{N_{\diamond}}\bigr)\herm{\mathbf{U}_{\diamond}} + \tfrac{1}{\gamma}\mathbf{U}_{\diamond}^{\bot}\mathbf{U}_{\diamond}^{\bot\mathrm{H}}.
			\end{equation}
			Every vector in $\complex^{K\nr}$ can be decomposed as $\mathbf{r} \triangleq \mathbf{r}_{\diamond} + \mathbf{r}_{\diamond}^{\bot}$, where $\mathbf{r}_{\diamond}\in\mathrm{V}_{\diamond}$ and $\mathbf{r}_{\diamond}^{\bot}\in\mathrm{V}_{\diamond}^{\bot}$.
			Each component is obtained by orthogonally projecting $\mathbf{r}$ onto the corresponding subspace, using projection matrices $\mathbf{P}_{\diamond}\triangleq\mathbf{U}_{\diamond}\herm{\mathbf{U}_{\diamond}}$ and $\mathbf{P}_{\diamond}^{\bot}\triangleq\mathbf{U}_{\diamond}^{\bot}\mathbf{U}_{\diamond}^{\bot\mathrm{H}}$, respectively.
			The \acrshort{pep} integral in terms of $\mathrm{V}_{\diamond}$ and $\mathrm{V}_{\diamond}^{\bot}$ is
			\begin{equation}
				\mathrm{P}_{a\to b} = \int_{\mathcal{P}'}\hspace{-.5em} \frac{\euler^{-\herm{\mathbf{r}_{a}}\mathbf{U}_a(\mathbf{\Lambda}_a+\frac{1}{\gamma}\id{N_a})^{-1}\herm{\mathbf{U}_a}\mathbf{r}_{a}}}{\pi^{N_a}\vert\mathbf{\Lambda}_a+\frac{1}{\gamma}\id{N_a}\vert}\frac{\euler^{-\gamma\lVert\mathbf{r}_a^{\bot}\rVert^2}}{(\pi/\gamma)^{N_{a}^{\bot}}}\der\mathbf{r}_a\der\mathbf{r}_a^{\bot}\label{eq:integral_split}.
			\end{equation}
			To obtain its limit as $\gamma\to\infty$, we apply a simple change of variable $\mathbf{t}_a^{\bot}\triangleq\sqrt{\gamma}\cdot\mathbf{r}_a^{\bot}$:
			\begin{equation}
				\mathrm{P}_{a\to b} = \int_{\mathcal{P}'}\hspace{-.5em} \frac{\euler^{-\herm{\mathbf{r}_{a}}\mathbf{U}_a(\mathbf{\Lambda}_a+\frac{1}{\gamma}\id{N_a})^{-1}\herm{\mathbf{U}_a}\mathbf{r}_{a}}}{\pi^{N_a}\vert\mathbf{\Lambda}_a+\frac{1}{\gamma}\id{N_a}\vert} \frac{\euler^{-\lVert\mathbf{t}_a^{\bot}\rVert^2}}{\pi^{N_{a}^{\bot}}}\der\mathbf{r}_a\der\mathbf{t}_a^{\bot}.\label{eq:pep2}
			\end{equation}
			The integrand of this expression converges pointwise to
			\begin{equation}
				\tfrac{1}{\pi^{N_a}\vert\mathbf{\Lambda}_a\vert}\euler^{-\herm{\mathbf{r}_{a}}\mathbf{U}_a\mathbf{\Lambda}_a^{-1}\herm{\mathbf{U}_a}\mathbf{r}_{a}}\cdot\tfrac{1}{\pi^{N_{a}^{\bot}}}\euler^{-\lVert\mathbf{t}_a^{\bot}\rVert^2}
			\end{equation}
			as $\gamma\to\infty$.
			This allows the use of Lebesgue's Dominated Convergence Theorem~\cite[Thm.~11.3.13]{Choudary2014} onto the limit~of~\eqref{eq:pep2}:
			\begin{equation}
				\smashoperator{\lim_{\gamma\to\infty}}\mathrm{P}_{a\to b} = \int_{\mathcal{P}_{\infty}'}\hspace{-1em} \frac{\euler^{-\herm{\mathbf{r}_{a}}\mathbf{U}_a\mathbf{\Lambda}_a^{-1}\herm{\mathbf{U}_a}\mathbf{r}_{a}}}{\pi^{N_a}\vert\mathbf{\Lambda}_a\vert}\frac{\euler^{-\lVert\mathbf{t}_a^{\bot}\rVert^2}}{\pi^{N_{a}^{\bot}}}\der\mathbf{r}_a\der\mathbf{t}_a^{\bot},\label{eq:pep3}
			\end{equation}
			where $\mathcal{P}_{\infty}'\triangleq\lim_{\gamma\to\infty}\mathcal{P}'$.
			This region of integration has been developed in~\eqref{eq:region}, where we have defined
			\begin{equation}
				\kappa_{a,b}(\gamma)\triangleq \ln\lvert\mathbf{\Lambda}_a\rvert - \ln\lvert\mathbf{\Lambda}_b\rvert + (N_a-N_b)\ln\gamma.
			\end{equation}
			
			\begin{figure*}[!b]
				\normalsize
				\hrulefill
				\begin{equation}
					\mathcal{P}_{\infty}' = \Bigl\{ \mathbf{r}\in\complex^{K\nr} : \lVert\herm{\mathbf{U}_{b}}\mathbf{r}_a\rVert_{\mathbf{\Lambda}_{b}}^2 - \lVert\herm{\mathbf{U}_{a}}\mathbf{r}_a\rVert_{\mathbf{\Lambda}_{a}}^2 +
					\lVert\mathbf{U}_{b}^{\bot\mathrm{H}}\mathbf{t}_a^{\bot}\rVert^2 - \lVert\mathbf{t}_a^{\bot}\rVert^2 + \smashoperator{\lim_{\gamma\to\infty}}\gamma\lVert\mathbf{U}_{b}^{\bot\mathrm{H}}\mathbf{r}_a\rVert^2 +2\sqrt{\gamma}\Re\bigl\{\herm{\mathbf{r}_a}\mathbf{P}_{b}^{\bot}\mathbf{t}_a^{\bot}\bigr\}\leq \smashoperator{\lim_{\gamma\to\infty}}\kappa_{a,b}(\gamma) \Bigr\} \label{eq:region}
				\end{equation}
			\end{figure*}
			
			The high \acrshort{snr} limit of the \acrshort{pep} in~\eqref{eq:pep3} involves integrating a non-negative density function over $\mathcal{P}_{\infty}'$.
			Therefore, to evaluate its behavior, we must analyze the structure of such region under every configuration of $\mathrm{V}_a$ and $\mathrm{V}_b$.
			We can immediately notice that the \acrfull{lhs} of~\eqref{eq:region} will remain bounded as $\gamma\to\infty$ when $\mathbf{P}_b^{\bot}\mathbf{r}_a=\mathbf{0}$, \ie for all $\mathbf{r}\in\complex^{K\nr}$ such that their orthogonal projection onto $\mathrm{V}_a$ belongs to $\mathrm{V}_b$.
			This will only occur when $\mathrm{V}_a\subseteq\mathrm{V}_b$.
			Otherwise, it will diverge with $O(\gamma)$.
			
			Based on these observations, we can study three separate scenarios:
			\begin{enumerate}
				\item $\mathrm{V}_a\nsubseteq\mathrm{V}_b$:
				The \acrshort{lhs} of~\eqref{eq:region} diverges to $\infty$ as $\gamma\to\infty$ for every $\mathbf{r}\in\complex^{K\nr}$.
				The right-hand side can display different behaviors depending on $N_a$ and $N_b$:
				\begin{itemize}
					\item $N_a<N_b$: It diverges to $-\infty$.
					\item $N_a=N_b$: It remains bounded.
					\item $N_a>N_b$: It diverges to $\infty$ with $O(\ln\gamma)$, \ie slower than the \acrshort{lhs}.
				\end{itemize}
				There is no element of $\complex^{K\nr}$ that belongs to $\mathcal{P}_{\infty}'$ in any of the three possible cases.
				Therefore, $\mathcal{P}_{\infty}' = \{\emptyset\}$ and the \acrshort{pep} will vanish.
				\item $\mathrm{V}_a\subset\mathrm{V}_b$:
				The \acrshort{lhs} of~\eqref{eq:region} remains bounded as $\gamma\to\infty$.
				Since $N_a<N_b$, $\lim_{\gamma\to\infty}\kappa_{a,b}(\gamma)=-\infty$, so no element of $\complex^{K\nr}$ belongs to $\mathcal{P}_{\infty}'$ and the \acrshort{pep} will vanish.
				\item $\mathrm{V}_a\equiv\mathrm{V}_b$:
				Both sides of~\eqref{eq:region} remain bounded, since $N_a=N_b$.
				The region of integration reduces to
				\begin{multline}
					\mathcal{P}_{\infty}' = \bigl\{ \mathbf{r}\in\complex^{K\nr} : \lVert\herm{\mathbf{U}_{b}}\mathbf{r}_a\rVert_{\mathbf{\Lambda}_{b}}^2 - \lVert\herm{\mathbf{U}_{a}}\mathbf{r}_a\rVert_{\mathbf{\Lambda}_{a}}^2 \\
					\leq \ln\lvert\mathbf{\Lambda}_a\rvert - \ln\lvert\mathbf{\Lambda}_b\rvert \bigr\},
				\end{multline}
				which is delimited by a quadric in $\mathrm{V}_a$ and does not depend on $\mathbf{t}_a^{\bot}$.
				The asymptotic \acrshort{pep} in~\eqref{eq:pep3} is thus
				\begin{align}
					\smashoperator{\lim_{\gamma\to\infty}}\mathrm{P}_{a\to b} &= \int_{\mathcal{P}_{\infty}'}\hspace{-.75em} \frac{\euler^{-\lVert\herm{\mathbf{U}_a}\mathbf{r}_{a}\rVert_{\mathbf{\Lambda}_a}^2}}{\pi^{N_a}\vert\mathbf{\Lambda}_a\vert} \der\mathbf{r}_a \int_{\mathrm{V}_a^{\bot}}\hspace{-.5em} \frac{\euler^{-\lVert\mathbf{t}_a^{\bot}\rVert^2}}{\pi^{N_{a}^{\bot}}}\der\mathbf{t}_a^{\bot}\nonumber\\
					&= \int_{\mathcal{P}_{\infty}'}\hspace{-.75em} \frac{\euler^{-\lVert\herm{\mathbf{U}_a}\mathbf{r}_{a}\rVert_{\mathbf{\Lambda}_a}^2}}{\pi^{N_a}\vert\mathbf{\Lambda}_a\vert} \der\mathbf{r}_a>0.
				\end{align}
				Therefore, when $\mathrm{V}_a\equiv\mathrm{V}_b$, the \acrshort{pep} is positively lower-bounded.
			\end{enumerate}
			
			As proved in the previous analysis, the only configuration in which the \acrshort{pep} does not vanish as $\gamma\to\infty$ is when $\mathrm{V}_a\equiv\mathrm{V}_b$.
			Hence, error-free detection can be asymptotically achieved in the high \acrshort{snr} regime by using an alphabet $\mathcal{S}$ such that
			\begin{equation}
				\colsp(\mathbf{\Xi}_{a}) \neq \colsp(\mathbf{\Xi}_{b}) \iff \mathbf{S}_a\neq\mathbf{S}_b,\quad \forall \mathbf{S}_a,\mathbf{S}_b\in\mathcal{S}.
			\end{equation}
			From~\cite[Thm.~3.5.3]{Bernstein2018}, we know that $\colsp(\mathbf{\Xi}_{\diamond}) = \colsp(\overline{\mathbf{C}}_{\vecrand{z}}\vecdet{S}_{\diamond}\cov{\vecrand{h}})$.
			By definition, this column space is stated as
			\begin{equation}
				\colsp(\mathbf{\Xi}_{\diamond}) \triangleq \bigl\{\mathbf{u}\in\colsp\bigl(\overline{\mathbf{C}}_{\vecrand{z}}\bigr) : \vecdet{S}_{\diamond}\mathbf{v} = \mathbf{u},\, \forall \mathbf{v}\in\colsp\bigl(\cov{\vecrand{h}}\bigr)\bigr\}.
			\end{equation}
			Since both $\overline{\mathbf{C}}_{\vecrand{z}}$ and $\cov{\vecrand{h}}$ are rank-complete, their column spaces are $\colsp(\overline{\mathbf{C}}_{\vecrand{z}}) = \complex^{K\nr}$ and 
			$\colsp(\cov{\vecrand{h}}) = \complex^{\nt\nr}$.
			Therefore,
			\begin{equation}
				\colsp(\mathbf{\Xi}_{\diamond}) = \bigl\{\mathbf{u} : \vecdet{S}_{\diamond}\mathbf{v} = \mathbf{u}\bigr\} \equiv \colsp\bigl(\vecdet{S}_{\diamond}\bigr).
			\end{equation}
			Considering these derivations, and by construction of $\vecdet{S}_{\diamond}$, it is straightforward to see that
			\begin{equation}
				\colsp(\mathbf{\Xi}_{a}) \neq \colsp(\mathbf{\Xi}_{b}) \iff \colsp(\mathbf{S}_a) \neq \colsp(\mathbf{S}_b).
			\end{equation}
			This completes the proof.
		\end{IEEEproof}
		
		\subsection{Comments on Proposition~\ref{prop:subspace}}
		
			The previous result states that an alphabet will allow for \acrshort{asd} in the high \acrshort{snr} regime if and only if each codeword spans a different subspace.
			In broad strokes, this implies that the spectrum shape of $\cov{\diamond}$ becomes irrelevant at high \acrshort{snr} and only the signal geometry remains\footnote{This phenomenon is related to the \textit{estimator-correlator}, whose coefficients flatten as the \acrshort{snr} increases~\cite[Sec.~5.3]{Kay1998}.}.
			In a sense, the difference in covariance matrices required for unique identification (Lemma~\ref{lmm:ui}) is replaced by a difference in signal projection matrices.
			
			Proposition~\ref{prop:subspace} is a generalization of~\cite[Thm.~2]{VilaInsa2024a} for any $K$ and $\nt$.
			In that work, energy-based constellations were investigated for $\nt=K=1$.
			The existence of a high-\acrshort{snr} error floor was proved in such scenario if and only if $M>2$.
			The same conclusion can be reached from Proposition~\ref{prop:subspace}:
			\begin{itemize}
				\item When $M=2$, the transmitted symbols are $x_0=0$ and $x_1>0$.
				The null symbol spans a subspace of dimension 0, while $x_1$ spans one of maximum dimension at the receiver.
				By Proposition~\ref{prop:subspace}, this guarantees \acrshort{asd}.
				\item On the contrary, when $M>2$, at least a pair of symbols will have non-null energy, both spanning the full available space at the receiver.
				Since their \acrshort{pep} will be positively lower-bounded, the detection of such constellation will have an error floor at high \acrshort{snr}.
			\end{itemize}
			
			Another noteworthy implication of this result is how it relates to \textit{\acrfull{ustm}}~\cite{Hochwald2000}.
			It is known that, under isotropic Rayleigh fading, it achieves a vanishing gap from the channel capacity in the high \acrshort{snr} regime for some configurations of $K$, $\nt$ and $\nr$~\cite{Zheng2002}.
			Moreover, it reaches the optimal degrees of freedom of the channel in various other cases~\cite{Yang2013}.
			For $K\geq\nt$, each codeword in \acrshort{ustm} is constructed from a truncated unitary matrix~\cite{Ngo2022} (\ie $\herm{\mathbf{S}_i}\mathbf{S}_i=\id{\nt}$ and $\mathbf{S}_i\herm{\mathbf{S}_i}=\mathbf{P}_i$) and corresponds to a different point in the \textit{Grassmann manifold} $\mathcal{G}(\nt,\complex^{K})$~\cite{Zheng2002}.
			Therefore, the columns of each element in a \textit{\acrfull{gc}} span a different $\nt$-dimensional subspace in $\complex^{K}$.
			
			It is clear that these codewords satisfy condition~\eqref{eq:prop_condition_2} and yield \acrshort{asd} in the high \acrshort{snr} regime.
			However, Proposition~\ref{prop:subspace} hints at a more general class of constellations, of which \acrshort{gc}s are a special case such that each codeword spans a subspace of the same dimension.
			A \textit{subspace-based codebook} that would relax this constraint could be constructed as
			\begin{equation}
				\mathcal{S}_{\text{subspace}}\triangleq\bigcup_{n=0}^{\nt}\mathcal{S}_{n},
			\end{equation}
			where $\mathcal{S}_n$ is a \acrshort{gc} corresponding to $\mathcal{G}(n,\complex^K)$.
			Notice that $\mathcal{G}(0,\complex^K)$ will contain a single element at most (\ie the null codeword $\mathbf{0}_{K\times\nt}$).
			The same occurs with $\mathcal{G}(\nt,\complex^K)$ when $\nt=K$, in which the only possible codeword spans the full available $K$-dimensional space.
	
	\section{Concluding remarks}
	
		This work has explored essential aspects of noncoherent detection, by translating well-established results from detection theory to the context of noncoherent \acrshort{mimo} communications.
		In particular, necessary and sufficient conditions for \acrshort{asd} have been obtained in the large array and high \acrshort{snr} regimes.
		These results have provided new insights onto relevant metrics and signal structures under each scenario.
		
		On the one hand, Proposition~\ref{prop:nr} establishes a principle for testing whether the error probability of a system model suffers a fundamental limitation that cannot be overcome by increasing the number of receiving elements.
		A variety of channel profiles and codewords can be analyzed under this criterion.
		On the other hand, Proposition~\ref{prop:subspace} reveals an umbrella class of alphabets that achieve \acrshort{asd} at high \acrshort{snr}.
		The presented framework leads to further research, for example in the direction of deriving benchmarks and/or alternatives to \acrshort{gc}s, exhibiting potentially improved performance and detection complexity.

	\bibliographystyle{IEEEtran}
	\bibliography{IEEEabrv,refs}

\begin{thebibliography}{10}
\providecommand{\url}[1]{#1}
\csname url@samestyle\endcsname
\providecommand{\newblock}{\relax}
\providecommand{\bibinfo}[2]{#2}
\providecommand{\BIBentrySTDinterwordspacing}{\spaceskip=0pt\relax}
\providecommand{\BIBentryALTinterwordstretchfactor}{4}
\providecommand{\BIBentryALTinterwordspacing}{\spaceskip=\fontdimen2\font plus
\BIBentryALTinterwordstretchfactor\fontdimen3\font minus
  \fontdimen4\font\relax}
\providecommand{\BIBforeignlanguage}[2]{{%
\expandafter\ifx\csname l@#1\endcsname\relax
\typeout{** WARNING: IEEEtran.bst: No hyphenation pattern has been}%
\typeout{** loaded for the language `#1'. Using the pattern for}%
\typeout{** the default language instead.}%
\else
\language=\csname l@#1\endcsname
\fi
#2}}
\providecommand{\BIBdecl}{\relax}
\BIBdecl

\bibitem{Goldsmith2005}
A.~Goldsmith, \emph{Wireless Communications}.\hskip 1em plus 0.5em minus
  0.4em\relax Cambridge, U.K.: Cambridge Univ. Press, 2005.

\bibitem{Chowdhury2016}
M.~Chowdhury, A.~Manolakos, and A.~Goldsmith, ``Scaling laws for noncoherent
  energy-based communications in the {SIMO MAC},'' \emph{IEEE Trans. Inf.
  Theory}, vol.~62, no.~4, pp. 1980--1992, Apr. 2016.

\bibitem{Chafii2023}
M.~Chafii, L.~Bariah, S.~Muhaidat, and M.~Debbah, ``Twelve scientific
  challenges for {6G}: Rethinking the foundations of communications theory,''
  \emph{IEEE Commun. Surv. Tutor.}, vol.~25, no.~2, pp. 868--904, 2023.

\bibitem{Jing2016}
L.~Jing, E.~D. Carvalho, P.~Popovski, and A.~O. Martinez, ``Design and
  performance analysis of noncoherent detection systems with massive receiver
  arrays,'' \emph{IEEE Trans. Signal Process.}, vol.~64, no.~19, pp.
  5000--5010, Oct. 2016.

\bibitem{VilaInsa2024a}
M.~Vil\`a-Insa, A.~Mart\'i, J.~Riba, and M.~Lamarca, ``Quadratic detection in
  noncoherent massive {SIMO} systems over correlated channels,'' \emph{IEEE
  Trans. Wirel. Commun.}, vol.~23, no.~10, pp. 14\,259--14\,272, Oct. 2024.

\bibitem{Heath2018}
R.~W. Heath~Jr. and A.~Lozano, \emph{Foundations of MIMO Communication}.\hskip
  1em plus 0.5em minus 0.4em\relax Cambridge, U.K.: Cambridge Univ. Press,
  2018.

\bibitem{Pizzo2022}
A.~Pizzo, L.~Sanguinetti, and T.~L. Marzetta, ``Fourier plane-wave series
  expansion for holographic {MIMO} communications,'' \emph{IEEE Trans. Wirel.
  Commun.}, vol.~21, no.~9, pp. 6890--6905, Sep. 2022.

\bibitem{Kailath1975}
T.~Kailath and H.~Weinert, ``An {RKHS} approach to detection and estimation
  problems--{II}: {Gaussian} signal detection,'' \emph{IEEE Trans. Inf.
  Theory}, vol.~21, no.~1, pp. 15--23, Jan. 1975.

\bibitem{Bernstein2018}
D.~S. Bernstein, \emph{Scalar, Vector, and Matrix Mathematics}, 3rd~ed.\hskip
  1em plus 0.5em minus 0.4em\relax Princeton, NJ, USA: Princeton Univ. Press,
  2018.

\bibitem{Lapidoth2017}
A.~Lapidoth, \emph{A Foundation in Digital Communication}, 2nd~ed.\hskip 1em
  plus 0.5em minus 0.4em\relax Cambridge, U.K.: Cambridge Univ. Press, 2017.

\bibitem{PetreStoica1990}
P.~Stoica and A.~Nehorai, ``Performance study of conditional and unconditional
  direction-of-arrival estimation,'' \emph{{IEEE} Trans. Acoust., Speech,
  Signal Process.}, vol.~38, no.~10, pp. 1783--1795, Oct. 1990.

\bibitem{Ngo2022}
K.-H. Ngo, S.~Yang, M.~Guillaud, and A.~Decurninge, ``Joint constellation
  design for noncoherent {MIMO} multiple-access channels,'' \emph{IEEE Trans.
  Inf. Theory}, vol.~68, no.~11, pp. 7281--7305, Nov. 2022.

\bibitem{Hajek1998}
R.~B. Marie~Hu\u{s}kov\'a and V.~Dupac, \emph{Collected works of Jaroslav
  H\'ajek}, ser. Wiley Series in Probability and Statistics.\hskip 1em plus
  0.5em minus 0.4em\relax Chichester, U.K.: Wiley, 1998.

\bibitem{Wang2005}
Z.~Wang and B.~Vemuri, ``{DTI} segmentation using an information theoretic
  tensor dissimilarity measure,'' \emph{IEEE Trans. Med. Imaging}, vol.~24,
  no.~10, pp. 1267--1277, Oct. 2005.

\bibitem{Kennedy2013}
R.~A. Kennedy and P.~Sadeghi, \emph{Hilbert Space Methods in Signal
  Processing}.\hskip 1em plus 0.5em minus 0.4em\relax Cambridge, U.K.:
  Cambridge Univ. Press, 2013.

\bibitem{Kailath1998}
T.~Kailath and H.~Poor, ``Detection of stochastic processes,'' \emph{IEEE
  Trans. Inf. Theory}, vol.~44, no.~6, pp. 2230--2231, Oct. 1998.

\bibitem{Moakher2012}
M.~Moakher, ``Divergence measures and means of symmetric positive-definite
  matrices,'' in \emph{New Developments in the Visualization and Processing of
  Tensor Fields}, D.~H. Laidlaw and A.~Vilanova, Eds.\hskip 1em plus 0.5em
  minus 0.4em\relax Berlin, Heidelberg: Springer Berlin Heidelberg, 2012, pp.
  307--321.

\bibitem{Choudary2014}
A.~D.~R. Choudary and C.~P. Niculescu, \emph{Real Analysis on Intervals}.\hskip
  1em plus 0.5em minus 0.4em\relax New Delhi, India: Springer India, 2014.

\bibitem{Kay1998}
S.~Kay, \emph{Fundamentals of Statistical Signal Processing, Volume II:
  Detection Theory}.\hskip 1em plus 0.5em minus 0.4em\relax Upper Saddle River,
  NJ, USA: Prentice Hall, 1998.

\bibitem{Hochwald2000}
B.~Hochwald and T.~Marzetta, ``Unitary space-time modulation for
  multiple-antenna communications in {Rayleigh} flat fading,'' \emph{IEEE
  Trans. Inf. Theory}, vol.~46, no.~2, pp. 543--564, Mar. 2000.

\bibitem{Zheng2002}
L.~Zheng and D.~Tse, ``{Communication on the Grassmann manifold: a geometric
  approach to the noncoherent multiple-antenna channel},'' \emph{IEEE Trans.
  Inf. Theory}, vol.~48, no.~2, pp. 359--383, Feb. 2002.

\bibitem{Yang2013}
W.~Yang, G.~Durisi, and E.~Riegler, ``On the capacity of large-{MIMO}
  block-fading channels,'' \emph{IEEE J. Sel. Areas Commun.}, vol.~31, no.~2,
  pp. 117--132, Feb. 2013.

\end{thebibliography}
	
	\newpage
	\vfill

\end{document}